# The Chemical Instability of $Na_2IrO_3$ in Air

J. Krizan*, J.H. Roudebush, G. M. Fox and R.J. Cava

Department of Chemistry, Princeton University, Princeton NJ 08544

* Corresponding Author: jkrizan@princeton.edu

**Abstract**

We report that $Na_2IrO_3$, which has a layered honeycomb iridium oxide sublattice interleaved by Na planes, decomposes in laboratory air while maintaining the same basic crystal structure. The decomposition reaction was monitored by time-dependent powder x-ray diffraction under different ambient atmospheres, through which it was determined that it occurs only in the simultaneous presence of both $CO_2$ and $H_2O$. A hydrated sodium carbonate is the primary decomposition product along with altered $Na_2IrO_3$. The diffraction signature of the altered $Na_2IrO_3$ is quite similar to that of the pristine material, which makes the detection of decomposition difficult in a sample handled under ordinary laboratory conditions. The decomposed samples show a significantly decreased magnetic susceptibility and the disappearance of the low temperature antiferromagnetic transition considered to be characteristic of the phase. Samples that have never been exposed to air after synthesis display a previously unreported magnetic transition at 5K.

Highlights:

- Na2IrO3 decomposes rapidly in laboratory air.
- The decomposition requires the simultaneous presence of $CO_2$ and $H_2O$.
- Decomposition results in a dramatic change in the magnetic properties.
- Second 5K feature in magnetic susceptibility not previously reported.

Graphical Abstract:

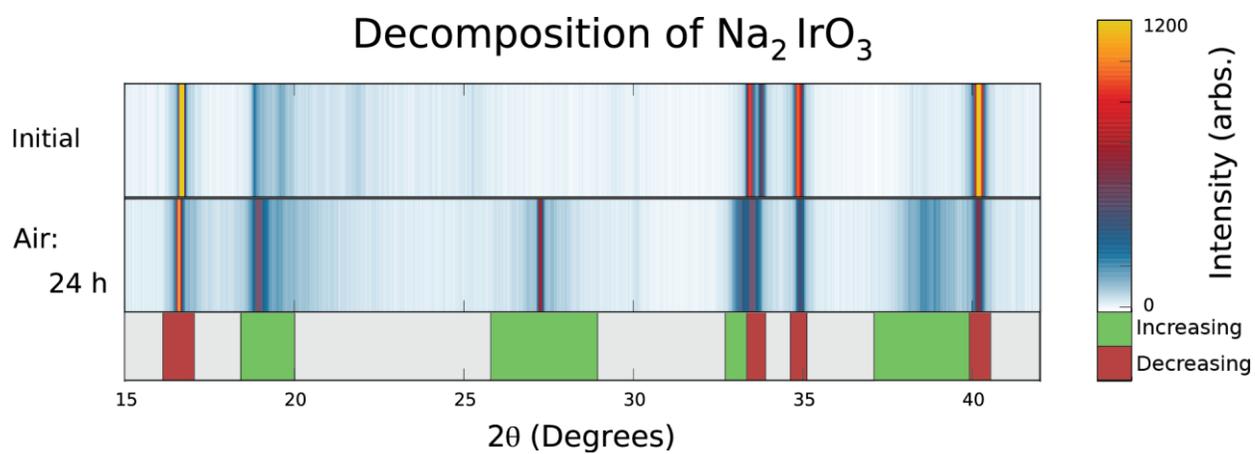

**Introduction**

Layered transition metal oxides have often been studied as a test bed of strongly correlated electronic phenomena. Recently, iridates have attracted attention in this regard because the spin-orbit interactions and electron correlations present are believed to be of comparable strength, which is expected to lead to non-trivial electronic ground states and exotic phenomena. Iridates have been looked to, for example, as experimental realizations of a spin liquid state and the quantum spin Hall Effect, as well as potential topological insulators[1–4]. $Na_2IrO_3$ has been of particular interest as a potential embodiment of "Kitaev-model physics," where $J = ½$ ions are arranged on a honeycomb lattice.[3,5,6]

$Na_2IrO_3$ is an ordered supercell variant of the layered α-$NaFeO_2$ structure type. In this structure type, $MO_6$ (M = metal ion) octahedra form a planar triangular lattice separated by a layer of alkali atoms. To better illustrate the correspondence, $Na_2IrO_3$ can be written as $Na_3NaIr_2O_6$: one third of the sites within the $MO_6$ layer are replaced by sodium in an ordered fashion, and the other two-thirds are $IrO_6$ octahedra that form a $J=1/2$ honeycomb lattice. In this magneto-structural configuration, Kitaev-model physics predicts a gapless spin liquid state.[3,5] However, at the time of this publication, all experimental reports on $Na_2IrO_3$ indicate the presence of antiferromagnetic ordering at ~ 15 K. Consequently there are extensive spectroscopic and theoretical studies that address the models' inconsistencies with the experimental data.[7,8] $Na_2IrO_3$ is also predicted to be a topological insulator (TI), where a bulk band gap exists simultaneously with topologically protected metallic surface states;[1] at present there are no reports of the experimental realization of this phenomenon in $Na_2IrO_3$. Recent theoretical literature suggests that changes in the interlayer spacing of $Na_2IrO_3$ may trigger a quantum phase transition to a TI.[1]

Despite the large number of materials-physics-based studies that theoretically and experimentally address $Na_2IrO_3$, a body of work on its chemistry and stability is lacking. Well established studies on $NaNiO_2$, $NaTiO_2$, and $NaCoO_2$, which also display the α-$NaFeO_2$ structure, have shown that they often contain sodium deficiencies and are sensitive to moisture and are the motivation for our investigation.[9–11] We report here that chemical instability of $Na_2IrO_3$ under normal laboratory conditions is indeed present, and significantly affects the magnetic properties

.

**Experimental**

Polycrystalline samples of $Na_2IrO_3$ were prepared from starting materials $Na_2CO_3$ and Ir metal (Alfa, 99.95%), weighed with a 5% Na excess to accommodate Na volatility during synthesis. The loose, well mixed and ground powder was placed in a covered crucible and heated in air at 800 °C for 48 hours and then oven cooled to 600 °C. Each sample was then removed directly from the oven at 600 °C and pumped into a Ar-filled glove box with less than 0.1 ppm of $O_2$ while still hot to prevent reaction with the ambient atmosphere.

Laboratory X-ray powder diffraction (PXRD) experiments were carried out on a Bruker D8 Focus diffractometer with Cu Kα radiation and a graphite diffracted beam monochromator. For the in situ gas flow experiments, a purpose-built apparatus with a Kapton shroud to bathe the sample in the atmospheric environment was installed in the diffractometer. Diffraction characterization of the air-exposed material was carried out under static laboratory air. Of the reference patterns, one (here referred to as "pristine $Na_2IrO_3$") was taken on minimal exposure to air, in an Ar filled sample chamber, and the dry $CO_2$ experiment was carried out in a $CO_2$ stream directly from the cylinder. Wet $O_2$, $N_2$ and $CO_2$ diffraction measurements were carried out by bubbling $O_2$, $N_2$ or $CO_2$ gas through a sealed container of deionized $H_2O$. Thirty minute scans, with half hour intervals between scans, were taken between 15° and 45° 2θ.

The FullProf software suite was used for the Rietveld refinements and profile fits.[12] Peak shapes were modeled with the Thompson-Cox-Hastings pseudo-Voight profile convoluted with axial divergence asymmetry. Background was modeled with a Chebychev Polynomial. An important structural characteristic of this compound, and very common among layered structure types, is the presence of stacking faults. This presence of stacking faults of the honeycomb lattice can be modeled on average by a refinement of the apparent Na/Ir site disorder; this disorder is not within the honeycomb lattice plane itself but rather reflects variations in the stacking of the honeycomb from one plane to the next.[13]

Magnetic susceptibility measurements of powder samples weighing between 70-90 mg were made upon heating in a 1 Tesla field after cooling to 2 K in zero field using a Quantum Design Physical Properties Measurement System (PPMS). The "pristine" $Na_2IrO_3$ sample was loaded into the measurement container in the glove box.

**Results and Discussion**

An X-ray diffraction pattern along with a structural fit and refinement of the pristine $Na_2IrO_3$ sample is shown in Figure 1. The observed cell parameters and refined structure are in agreement with previous reports; they are compared in Table 1. In comparing these cell parameters, it appears that the literature cell parameters for the *a* and *b* axes are slightly smaller, approximately six standard deviations different, from those reported here; these small differences may be attributed either to subtleties in the experimental procedure or to the fact that under normal circumstances $Na_2IrO_3$ is handled in air.

The refined positional parameters from the pristine sample are presented in Table 2. The only structural parameters refined were those of the Na/Ir sites and the site mixing; the oxygen positions were not refined and were fixed at the previously reported positions due to the fact that laboratory XRD is not sensitive enough to detect subtle differences in oxygen positions. Comparing the extent of site mixing seen in the literature to that of the pristine sample, an estimate of the stacking faults present, shows that they are the same within error. Looking closer at the structure, we find that although three distinct Ir-Ir bond lengths are allowed by symmetry, they are in fact, due to the stacking faults, indistinguishable by powder diffraction. Due to the implications that a non-isometric honeycomb lattice might have on the magnetism, further consideration of this issue, employing a local structural probe, may be of future interest. While the refinements here are carried out in the space group C 2/c, the C 2/m space group also proposed for this material[8,14] presents an identical fit in the powder refinement. C 2/c was chosen for easy comparison to the literature unit cell and magnetism.[15] These two models are very similar, and the choice of space group is outside the scope of this work.

After establishing the purity of our pristine sample and its nominal (but not exact) equivalence to the widely-studied $Na_2IrO_3$ material, PXRD scans of the pristine sample exposed to ambient laboratory air were taken over a 24 hour time period in 30 min intervals. The results are shown in Figure 2. A steady change in the diffraction pattern is observed; the primary features of the change are a decrease in intensity of the peak at 16.7° 2θ, and an accompanying increase in intensity of the peak at 19.1° 2θ. To identify the cause of this change, PXRD measurements at ambient temperature in air, $CO_2$, $N_2/H_2O$, $O_2/H_2O$, and $CO_2/H_2O$ environments were performed. The results are shown in Figure 3. When $CO_2$, wet $O_2$, or $N_2$ gas was flowed over the sample, there were no significant changes in the diffraction pattern over the 12 hour period of the test. Due to the insensitivity of the compound to $CO_2$, $H_2O$ and $O_2$ alone, we concluded that that the instability of $Na_2IrO_3$ in air is due to the simultaneous presence of both $CO_2$ and $H_2O$ in air. Confirming this conclusion, exposing the sample to both $CO_2$ and $H_2O$ at ambient temperature at the same time, in higher concentration than is present in ambient air, resulted in its rapid degradation. The PXRD pattern from this rapidly degraded sample is the same as that seen in the sample exposed to air for a longer time. The observed behavior therefore indicates that there is a symbiotic mechanism involving both $CO_2$ and $H_2O$ that causes the air-instability of $Na_2IrO_3$.

The PXRD pattern and LeBail fit of the fully decomposed sample are shown in Figure 4. All of the peaks from the decomposed phase are allowed in the same monoclinic space group as, and with the same approximate unit cell of, the pristine $Na_2IrO_3$. The largest peak intensity is found for the reflection at 19.1° 2θ. This peak is coincident with the (110) reflection in the pristine sample. Because the pristine and decomposed sample share many reflections with very close 2θ positions, it is difficult to identify the presence of partial decomposition in a $Na_2IrO_3$ sample. Small peaks in the decomposed sample that are not indexed by the major phase were identified as being due to the presence of thermonatrite, $Na_2CO_3 \cdot H_2O$, as shown in Figure 4. The observation of $Na_2CO_3 \cdot H_2O$ in the decomposed sample further corroborates the identification of simultaneous exposure to both $CO_2$ + $H_2O$ as the cause of the decomposition. The decomposed $Na_2IrO_3$-like product shows a significant decrease in the $c$-axis spacing, by 0.192(3) Å. Similarly, the $a$ and $b$ axes contract, by 0.089(5) Å and 0.0725(9) Å respectively, changes that are well beyond experimental uncertainty (Table 1). Despite the broader diffraction peaks of the decomposed phase, an indication that its crystallinity is worse than that of the pristine material (not a surprise for a chemical reaction in the solid phase at ambient temperature), the observed contraction of the $c$-axis is significant, as the difference is over 40 times the standard deviation of the fits. β was fixed at the value determined from the refinement of the pristine sample to facilitate this comparison. This difference is 8 times higher than the sample-to sample variation for multiple preparations of $Na_2IrO_3$. Again looking at the literature refinement[15], the slightly smaller lattice parameters observed in that case could suggest the early stages of decomposition in that sample.[15] Our crystallographic evidence, coupled with the low temperature and mild conditions of the decomposition, indicates that the overall framework of $MO_6$ octahedra is unchanged during the decomposition. We hypothesize that the rigid bonding in the iridium honeycomb holds the layer structurally intact, and that the degradation is due to the deintercalation of a fraction of the sodium within the layers to form the hydrated sodium carbonate. This kind of reactivity is not unexpected in a material with such high alkali content where the alkali ions are found in layers and therefore are likely to have high ionic mobilities.[16]

Recently the sensitivity of $Ir^{4+}$ oxides to $CO_2$ + $H_2O$ has been characterized in a study on $IrO_2$ as a pH sensor.[17] The proposed reaction mechanism suggests that in $IrO_2$, $Ir^{4+}$ can be reduced to $Ir^{3+}$ in the presence of $CO_2$ and $H_2O$. If a similar mechanism is acting in the present case of $Na_2IrO_3$, it would result in the production of $Na_2CO_3 \cdot H_2O$ (as observed in the decomposed sample) by the following reaction:

[1]     $2Na_2Ir^{+4}O_3 + 3H_2O + CO_2 \leftrightarrow 2NaIr^{+3}O(OH)_2 + Na_2CO_3 \cdot H_2O + \frac{1}{2} O_2$

On the other hand, the $Na_2CO_3 \cdot H_2O$ formed might equally indicate an oxidative decomposition. The resulting $Na_{2-x}IrO_3$ would necessarily have an admixture of $Ir^{5+}$ in the ordinarily $Ir^{4+}$ lattice. A possible reaction for the oxidative reaction is given below. This mechanism requires additional oxygen, however, and thus while it is a possibility in laboratory air, it is less likely in our experiments in the $CO_2/H_2O$ mixture, where any oxygen present would likely have to be adventitious:

[2]     $2(Na_2Ir^{+4}O_3) + H_2O + CO_2 + \frac{1}{2} O_2 \leftrightarrow 2NaIr^{+5}O_3 + Na_2CO_3 \cdot H_2O$.

This instability of $Na_2IrO_3$ in $CO_2 + H_2O$ containing environments such as laboratory air is important because the magnetic properties of this material are the subject of much current research. The magnetic properties of a pristine sample of $Na_2IrO_3$ are compared to the same sample after prolonged exposure to air in Figure 5. The pristine material displays a feature characteristic of an antiferromagnetic transition at ~15 K, which has been identified as the intrinsic magnetic ordering temperature.[18] In addition to this transition, however, we observe a distinct second transition at 5 K, which has not been previously reported for samples handled in a normal fashion. At this time we are not able to comment on its nature, but note that a similar transition has been suggested in $Li_2IrO_3$.[5] The decomposed sample, on the other hand, lacks any magnetic transition and shows a dramatically decreased magnetic susceptibility in the paramagnetic regime.

Curie-Weiss analysis of the magnetic data for the pristine sample indicates an effective moment ($\mu_{EFF}$) of 2.0(1) $\mu_B$/Ir, larger than the expected spin-only moment for an $s = ½$ ion, an indication that spin orbit coupling makes a significant contribution to the magnetism of this sample. The Weiss temperature ($\theta$) is -159(3) K, indicating strong antiferromagnetic interactions. Both of these values are larger in magnitude than those found in the literature.[15] We make two primary inferences from our data: first that the magnetic properties of "$Na_2IrO_3$" are critically dependent on exposure to air, and second, that the literature sample[15] may be partially decomposed as evidenced by subtleties of the decomposition process shown in Figure 2. A more rigorous study of the differences between samples is needed to unravel these complexities. Most clearly, the decomposed sample, which, it should be recalled, has some kind of alteration of the basic crystal structure of the pristine phase, shows a marked decrease in magnetic susceptibility. The Curie-Weiss analysis of the magnetic data shows a decrease in effective moment to 0.27(1) $\mu_B$/Ir and a Weiss temperature of -2.93(5) K. These values are given in Table 3, where they are compared to the values reported for the pristine and literature values. The decrease in the effective moment indicates a change in the oxidation state of $Ir^{4+}$ ($d^5$, $J = ½$) to either non-magnetic $Ir^{5+}$ ($d^4$) or $Ir^{3+}$ ($d^6$), consistent with the proposed decomposition mechanisms. The decrease in Weiss constant from -159(3) to -2.93(5) K indicates a significant decrease in antiferromagnetic coupling, which is consistent with the dramatically decreased magnetic moment.

**Conclusion**

This work illustrates the high sensitivity of the structure and magnetic properties of $Na_2IrO_3$ to air exposure, specifically to the combination of $CO_2$ and $H_2O$ found in air. Our data indicates that decomposition leads to the removal of Na from between $MO_6$ layers and the coincident formation of a Na deficient $Na_2IrO_3$-related phase and $Na_2CO_3 \cdot H_2O$. This mechanism suggests that the challenges posed by sodium non-stoichiometry found in $NaMO_2$ (M = Ti, Co, Ni) are also present in $Na_2IrO_3$, and that further research on the effects of synthetic procedure and handling on observed magnetic properties is warranted. Magnetic susceptibility measurements confirm a change in iridium oxidation state as a result of the decomposition through a significant decrease in effective moment and Weiss constant in the decomposed sample. Given the intense research being conducted on the properties of this material, it is important for researchers to be cognizant of its instability in air, so as to be able to take appropriate handling precautions. Further, assessment of the intrinsic properties of the stoichiometric compound should be approached with caution.

**Acknowledgements:**

This research was supported by the U. S. Department of Energy, Division of Basic Energy Sciences, Grant DE-FG02-08ER46544.

**Tables and Figures**

Table 1: Lattice parameters

| | a (Å) | b (Å) | c (Å) | β (°) | Mixing*(%) |
|---|---|---|---|---|---|
| Literature Report[15] | 5.4198(5) | 9.3693(3) | 10.7724(7) | 99.568(23) | 14.4(2)% |
| Pristine** | 5.4231(5) | 9.3917(8) | 10.7706(9) | 99.497(7) | 17(6)% |
| Decomposed*** | 5.4142(2) | 9.3192(9) | 10.579(3) | 99.490 | |
| Difference (decomposed – pristine) | 0.0089(5) | 0.0725(9) | 0.192(3) | | |

\* Estimation of stacking faults, not actual Na/Ir disorder on honeycomb layer.

\*\* Standard deviations from the fit multiplied by 3 to account for correlated residuals.[19]

\*\*\* Standard deviations from the fit multiplied by 4 to account for correlated residuals.[19]

Table 2: Structural parameters for $Na_2IrO_3$

| Atom | Site | x | y | z | Occ. | Cell Parameters | | |
|------|------|---|---|---|------|---|---|---|
| Ir1  | 8f   | 0.266(3) | 0.0816(12) | 0.00033 | 0.83(6) | a=5.4231(5) Å | b=9.3917(8) Å | c=10.7706(13) Å |
| Na1  |      |          |            |         | 0.17(6) | α=90° | β=99.497(7)° | γ=90° |
| Na2  | 4d   | 0.75     | 0.25       | 0       | 0.70(3) | Space Group: | | Temperature: |
| Ir2  |      |          |            |         | 0.30(3) | C 2/c | | 300 K |
| Na3  | 4e   | 0.5      | 0.580(15)  | 0.25    | 1 | | | |
| Na4  | 4e   | 0        | 0.469(15)  | 0.25    | 1 | Refinement Parameters | | |
| Na5  | 4e   | 0        | 0.75       | 0.25    | 1 | $R_p$ | $R_{wp}$ | $R_{exp}$ $\chi^2$ |
| O1   | 8f   | 0.3795   | 0.2777     | 0.8998  | 1 | 15.3 | 20.4 | 14.31 2.04 |
| O2   | 8f   | 0.5966   | 0.0513     | 0.12128 | 1 | | | |
| O3   | 8f   | 0.6579   | 0.413      | 0.1086  | 1 | | | |

Table 3: Magnetic Properties

| Sample | $\chi_0$ (cm$^3$/mol) | $\mu_{EFF}$ ($\mu$B/Ir) | $\theta$ (K) |
|---|---|---|---|
| Literature[15] | 3.0 x 10$^{-5}$ | 1.81(2) | -116(3) |
| Pristine | 2.5 x 10$^{-4}$ | 2.0(1) | -159(3) |
| Decomposed | 4.7 x 10$^{-4}$ | 0.27(1) | -2.93(5) |

**Figure 1:** Rietveld refinement of the structure of pristine Na$_2$IrO$_3$ in an Ar-filled sample chamber.

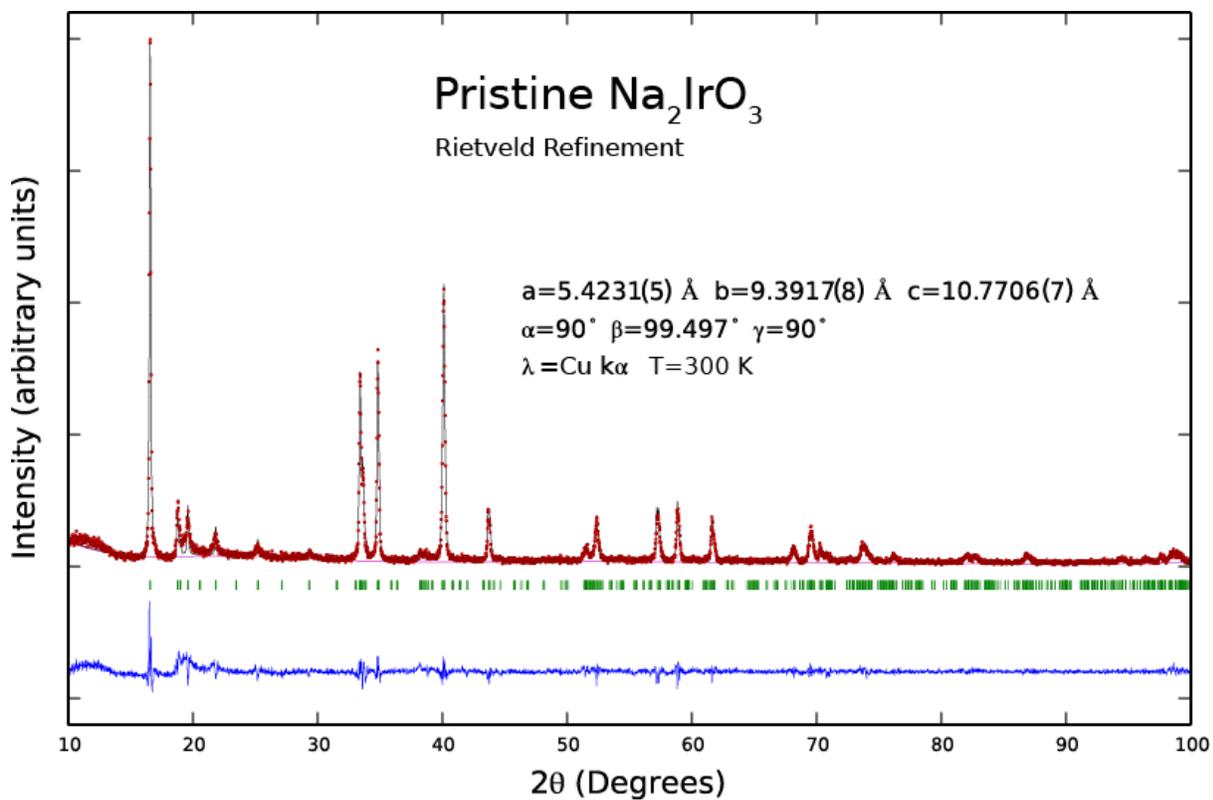

**Figure 2:** Decomposition of Na$_2$IrO$_3$ in air over the course of 24 hours. Half hour PXRD scans were taken with half hour intervals between each scan. Germanium powder was used as an internal standard (sharp peak at approx. 27° 2θ) to normalize intensities. The main reflections associated with pristine Na$_2$IrO$_3$ decrease in intensity while a number of broad regions of intensity and shoulders on existing peaks, characteristic of the decomposition, increase in intensity. Areas indicative of the decomposed product are at the 2θ values 19°, 27°, 33°, and 38.5° and are highlighted in the figure.

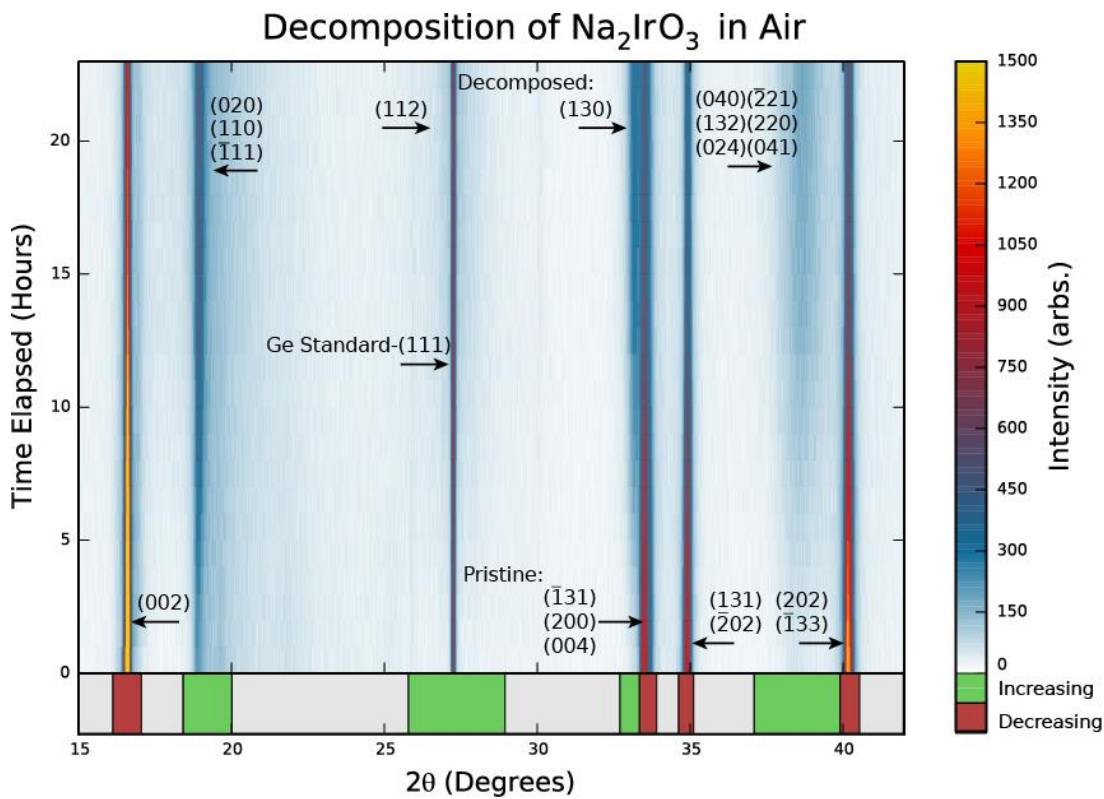

**Figure 3:** Comparison of PXRD patterns indicating the behavior of $Na_2IrO_3$ in different environments. Wet $N_2$, wet $O_2$, and dry $CO_2$ do not substantially affect the crystallinity of the pristine sample at room temperature over the course of 12 hours. Exposure to wet $CO_2$ results in a rapid degradation of the sample that is comparable to the decomposition seen in Air after 24 hours. In the air sample, the sharp feature at 27° 2θ is from the Ge internal standard. Areas indicative of the decomposition are highlighted in the figure.

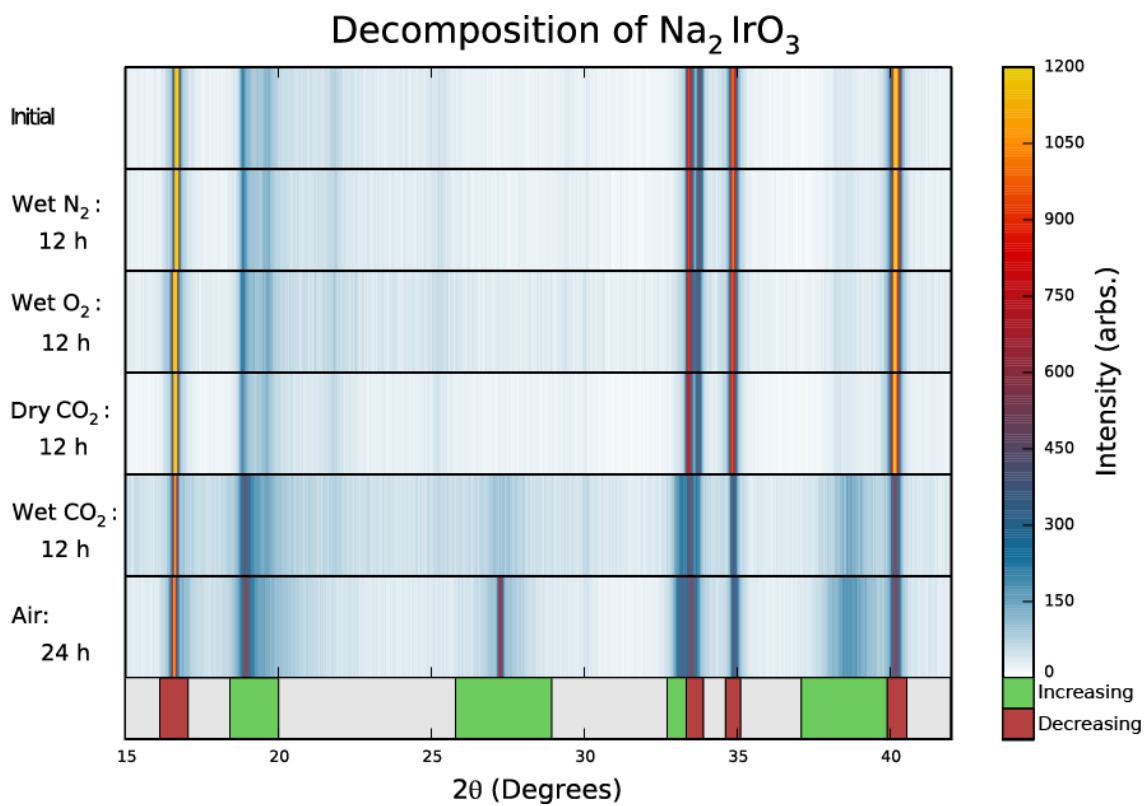

Figure 4: LeBail Profile fit to the decomposed Na$_2$IrO3 sample with the peaks from the Na$_2$CO$_3$•H$_2$O second phase marked. The insert shows the most prominent peaks (arrows) for Na$_2$CO$_3$•H$_2$O.

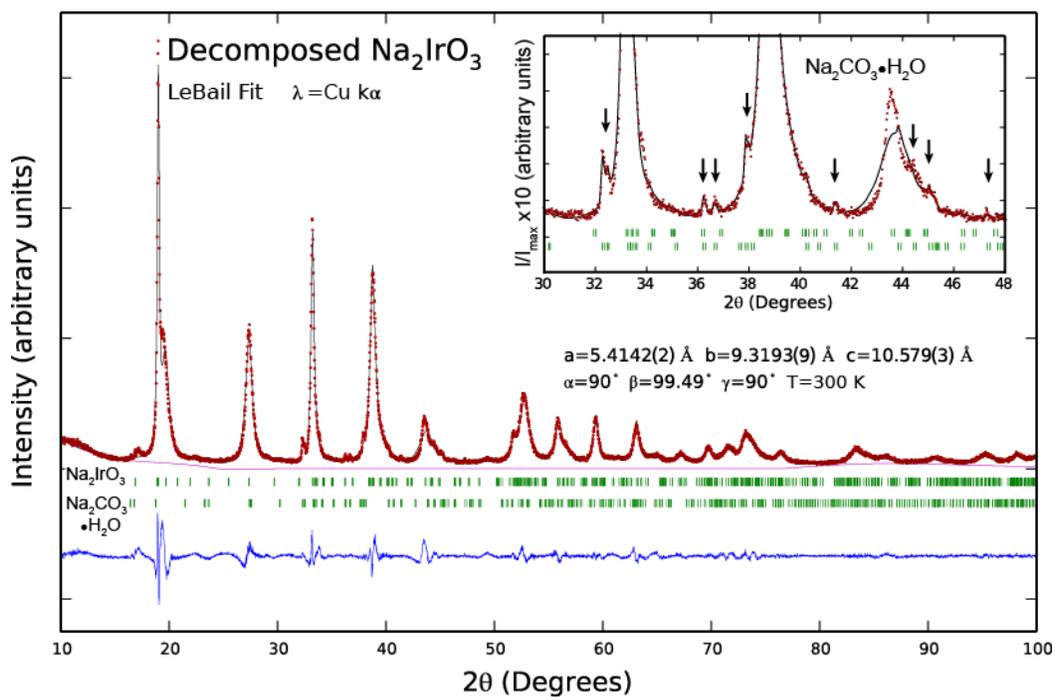

Figure 5: Temperature dependent magnetic susceptibility of pristine $Na_2IrO_3$ and the same sample exposed to air for an extended period, labeled as the "decomposed" sample. The insert shows the temperature dependence of the inverse susceptibility of the decomposed sample.

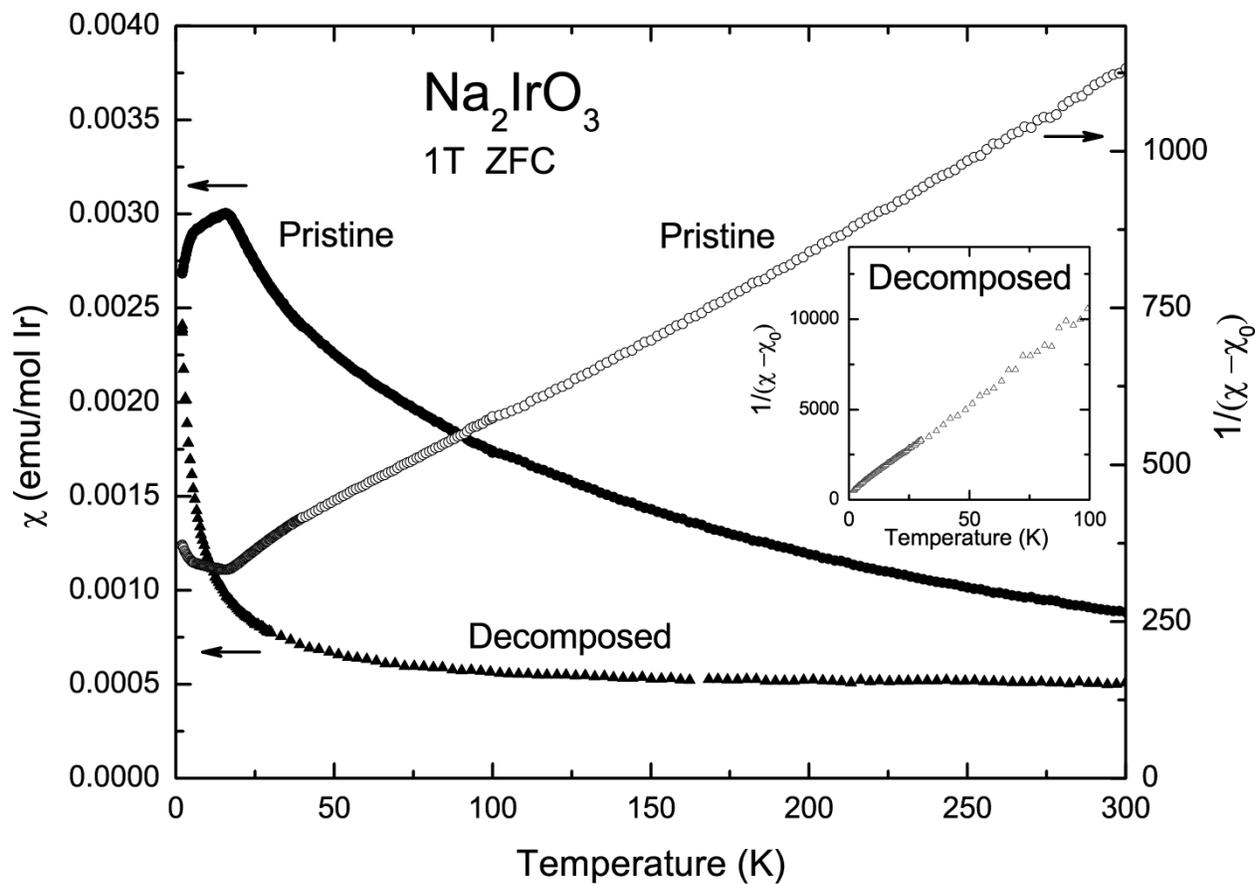